\documentclass{PoS}

\newcommand{\be}{ \begin{equation}}
\newcommand{\ee}{ \end{equation}}
\newcommand{\ba}{ \begin{array}}
\newcommand{\ea}{ \end{array}}
\newcommand{\bea}{ \begin{eqnarray}}
\newcommand{\eea}{ \end{eqnarray}}
\newcommand{\pref}[1]{(\ref{#1})}

\newcommand{\ca}{\overline{c}_{{\rm A}}}
\newcommand{\cv}{\overline{c}_{{\rm V}}}

\title{The vector and axial vector current in Wilson ChPT}

\ShortTitle{The vector and axial vector current in Wilson ChPT}

\author{Sinya Aoki\\
Graduate School of Pure and Applied Sciences\\
University of Tsukuba\\
Tsukuba, Ibaraki  305-8571\\
Japan\\
\vspace{-0.3cm}

Riken BNL Research Center\\
Brookhaven National Laboratory\\
Upton, NY 11973\\
USA\\
E-mail: \email{saoki@het.ph.tsukuba.ac.jp}
}
\author{\speaker{Oliver B\"ar}\\
        Institute of Physics\\
        Humboldt University Berlin\\
	12489 Berlin\\
	Germany\\
	E-mail: \email{obaer@physik.hu-berlin.de}
}

\abstract{We construct the vector and axial vector currents in Wilson Chiral Perturbation Theory (WChPT), the low-energy effective theory for lattice QCD with Wilson fermions.

Our construction is slightly different compared to ChPT in continuum QCD, where the currents are essentially the (partially) conserved currents associated with the chiral symmetries. In WChPT, due to explicit chiral symmetry breaking at non-zero lattice spacing, there appear O($a$) terms in the expressions for the currents which do not stem from the effective action. In addition, the finite renormalization of the currents needs to be taken into account in order to properly match the currents of the effective theory. 

As an illustration we compute $f_{\pi}$ to one loop with the renormalized axial vector current for a particular renormalization condition. It turns out that for this particular condition some of the O($a$) corrections are taken care of by the renormalization.
}

\FullConference{The XXV International Symposium on Lattice Field Theory\\
		 July 30 - August 4 2007\\
		 Regensburg, Germany}

\begin{document}

\section{Introduction}
Chiral perturbation theory (ChPT) is an important tool for Lattice QCD. It provides analytic guidance for the chiral extrapolation of the lattice data obtained at quark masses heavier than in nature. 

Continuum ChPT \cite{Gasser:1983yg,Gasser:1984gg} is based on the symmetries and the particular symmetry breaking of continuum QCD. Sharpe and Singleton have shown \cite{Sharpe:1998xm} how to construct ChPT for Lattice QCD with Wilson fermions \cite{Wilson:1974sk}, taking into account the explicit chiral symmetry breaking of Wilson's fermion discretization. The resulting low-energy effective theory, often called Wilson chiral perturbation theory (WChPT), is a double expansion in the quark mass and the lattice spacing, the two parameters of explicit chiral symmetry breaking.

In the following we discuss how the vector and axial vector currents are constructed in WChPT. 
Our construction of these currents is slightly different  compared to continuum ChPT. In order to illustrate the differences consider continuum QCD with two massless flavours. The QCD Lagrangian is then invariant under chiral transformations in $SU(2)_{R}\otimes SU(2)_{L}$. Consequently, there exist conserved non-singlet vector and axial vector currents,
\bea\label{CurrentsCont}
V_{\mu}^a & = & \overline{\psi}\gamma_{\mu}\frac{\sigma^a}{2}\psi\,,\qquad A_{\mu}^a\, = \,\overline{\psi}\gamma_{\mu}\gamma_5\frac{\sigma^a}{2}\psi\,.
\eea
In case the two flavours are massive but degenerate, the vector current is still conserved.
The conserved currents in \pref{CurrentsCont} are the Noether currents associated with the chiral symmetries and satisfy various chiral Ward identities (``current algebra'').

The symmetry $SU(2)_{R}\otimes SU(2)_{L}$ is spontaneously broken to its vector subgroup and the pions are the (pseudo) Goldstone bosons associated with this spontaneous symmetry breaking. At sufficiently low energies the pions are described by ChPT with the effective Lagrangian \cite{Gasser:1983yg,Gasser:1984gg}
\bea\label{ContLag} 
{\cal L}_{\rm eff}& =&  \frac{{f^2 }}{4}\left\langle \partial_{\mu} \Sigma (\partial_{\mu} \Sigma)^\dagger
\right\rangle + 
\frac{{f^2}B}{2} m\langle{\Sigma^{\dagger} + \Sigma}\rangle + \ldots
\eea
The field $\Sigma$ contains the pion fields in the usual way and the ellipsis denotes higher order terms. The symmetries of QCD constrain the form of the the chiral Lagrangian, and ${\cal L}_{\rm eff}$ essentially contains all terms compatible with the symmetries of QCD. At leading order in a derivative and mass expansion one obtains the two terms in \pref{ContLag}. 

Once the effective Lagrangian has been constructed the vector and axial vector can be derived from it. At leading order the (partially) conserved currents are identical to the Noether currents associated with the chiral symmetries
\bea\label{CurrentsChPT}
V_{\mu,{\rm ChPT}}^{a} & = &  \frac{f^{2}}{4} \langle\sigma^{a}(\Sigma^{\dagger}\partial_{\mu}\Sigma + \Sigma\partial_{\mu}\Sigma^{\dagger})\rangle\,,\qquad A_{\mu,{\rm ChPT}}^{a} \, = \, \frac{f^{2}}{4} \langle\sigma^{a}(\Sigma^{\dagger}\partial_{\mu}\Sigma - \Sigma\partial_{\mu}\Sigma^{\dagger})\rangle\,.
\eea

Now consider lattice QCD with two degenerate flavours of Wilson fermions. There exists a conserved vector current, but no partially conserved axial vector current due to the explicit chiral symmetry breaking by the Wilson term. Even though a conserved vector current exists, it is often not used in practice. The local, non-conserved vector current is employed instead, even though it requires the computation of a renormalization constant $Z_{V}$. Also the local axial vector current needs a non-trivial renormalization constant $Z_{A}$. One way to fix the renormalization constants is by imposing the chiral Ward identities and non-vanishing lattice spacing \cite{Karsten:1980wd,Bochicchio:1985xa,Maiani:1986yj}.

The explicit breaking of chiral symmetry and the need for renormalizing the currents raises the question how to construct the effective currents in Wilson ChPT. The ``Noether link'' does not hold anymore, even at leading order. Also the renormalization of the currents has to be taken into account for a proper matching of the effective theory to the fundamental lattice theory. 

In the following we present our results for the construction and matching for the two flavour theory. 
Details can be found in a forthcoming publication \cite{AB}.

\section{The currents in WChPT}
Starting point for our construction is the Symanzik effective theory \cite{Symanzik:1983dc} for Lattice QCD with Wilson fermions. Based on the locality and the symmetries of the lattice theory a dimensional analysis yields the following effective action \cite{Sheikholeslami:1985ij}:
\bea\label{SymAction}
S_{\rm Sym} & =&  S_{\rm QCD} + {a}\, \overline{c}_{sw}  \int d^{4}x\, \overline{\psi}(x) i \sigma_{\mu\nu} F_{\mu\nu}(x)\psi(x) + {\rm O}({a^2})\,.
\eea
This expression is essentially the most general Lorentz scalar compatible with the symmetries. 
Note, however,  that the equations of motion have been used in order to reduce the number of terms, which is legitimate as long as we are only interested in on-shell quantities.

The effective currents corresponding to the local lattice currents are obtained in the same way.
The most general vector and axial vector current compatible with locality and the symmetries read \cite{Luscher:1996sc,Sint:1997jx}:
\bea
V_{\mu,{\rm Sym, Loc}}^{a}  & = & V_{\mu,{\rm QCD}}^{a} + {a}\, \overline{c}_V \partial_{\nu}[\overline{\psi} \sigma_{\mu\nu} T^{a}\psi] + {\rm O}({a^{2}})\,,\label{SymVCurrent}\\
A_{\mu,{\rm Sym, Loc}}^{a} & = & A_{\mu,{\rm QCD}}^{a} + {a}\, \overline{c}_A \partial_{\mu}[\overline{\psi} \gamma_5 T^{a}\psi] + {\rm O}({a^{2}})\,.\label{SymACurrent}
\eea
Note that these currents are not the Noether currents obtained from the effective action \pref{SymAction}. The O($a$) corrections on the right hand side of \pref{SymVCurrent}, \pref{SymACurrent} are not directly related to the Pauli term in the action. 

In order to obtain the effective current for the conserved lattice vector current we have to impose current conservation on the vector current in \pref{SymVCurrent}. However, due to the tensor structure of the O($a$) correction the vector current in \pref{SymVCurrent} is already conserved through O($a$). Violations of current conservation will probably appear at O($a^{2}$). Nevertheless, at least through O($a$) the conserved and local vector current map onto the same Symanzik current, provided the latter includes the $Z$ factor.

Next we map the Symanzik currents to the chiral effective theory. Guiding principle is again that the chiral effective currents transform in the same way under symmetry transformations (parity, charge conjugation, chiral symmetry) as the currents in the Symanzik theory. Notice that the O($a$) corrections \pref{SymVCurrent}, \pref{SymACurrent} do not transform in the same way as the leading term because chiral symmetry is explicitly broken. For example, the vector current is the sum of the right and left handed currents, while the O($a$) tensor term couples right and left handed fields.
By promoting the coefficient $\overline{c}_V$ to a spurion field with the appropriate transformation law,  
the O($a$) correction can be made transform as the leading part, guarantying that the entire Symanzik vector current transforms in the same way as the QCD vector current. The same procedure is applied to the axial vector too. The chiral effective currents are subsequently obtained as the most general expressions transforming as vector and axial vector currents. Following this procedure we obtain
\bea
V_{\mu,{\rm WChPT}}^{a}  & = & V_{\mu,{\rm ChPT}}^{a}\left(1 + \frac{4}{f^{2}}a\,{W}_{V} \langle\Sigma + \Sigma^{\dagger}\rangle\right) + \ldots\,, \label{WChPTVC}\\
A_{\mu,{\rm WChPT}}^{a} & = & A_{\mu,{\rm ChPT}}^{a}\left(1 + \frac{4}{f^{2}} a\,{W}_{A,1}\langle\Sigma + \Sigma^{\dagger}\rangle\right) + 
2{a}\,{W}_{A,2}\partial_{\mu} \langle \sigma^{a}(\Sigma -\Sigma^{\dagger})\rangle +\ldots\,. \label{WChPTAC}
\eea 
Setting the the lattice spacing to zero we recover --by construction-- the continuum results
in \pref{CurrentsChPT}. The additional terms capture the leading linear O($a$) corrections, parameterized by unknown low-energy coefficients $W_{V},W_{A,1},W_{A,2}$ (we have absorbed the parameters $\cv,\ca$ in them since these coefficients are unknown too).  The ellipses stand for corrections of O$(am,ap^{2},a^{2}$) etc.

So far the effective currents are just arbitrary vector and axial vector currents. In case of the vector current we still have to impose current conservation, since the Symanzik vector currents are conserved at this order. Calculating $\partial_{\mu} V_{\mu,{\rm WChPT}}^{a} $ and taking into account the equation of motion we find the condition 
\bea\label{CondVCCons}
\partial_{\mu} V_{\mu,{\rm WChPT}}^{a} & = & 0 \quad \Leftrightarrow \quad W_{V}  = W_{45} \,,
\eea 
where  $W_{45} = W_{4} + W_{5}/2$ is a low-energy coefficient present in the chiral effective action \cite{Rupak:2002sm,Bar:2003mh,Sharpe:2004ny}. For the axial vector current there is no obvious additional condition which might relate $W_{A,1}$ or $W_{A,2}$ to the coefficients in the action, so we work with the axial current in \pref{WChPTAC}. 
 
The currents we constructed above are just the leading order ChPT currents modified by the first O($a$) correction. Higher order terms can be constructed in the same fashion. Among these terms are the standard NLO terms of continuum ChPT, of course. In addition, correction terms of O($ap^{2}, am)$ will also be present. Many terms contribute at this order and it is  straightforward but tedious  to find all possible terms contributing to the currents.

We remark that a somewhat different approach to the construction of the effective currents has been taken previously in Refs.\ \cite{Sharpe:2004ny,SharpeNara}, and it remains to be
clarified the extent to which the two approaches agree.

\section{Renormalization}

Since the local lattice currents are not conserved they require a finite renormalization. One introduces renormalized currents by
\bea\label{RenCurrents}
V_{\mu,{\rm ren}}^a & = & Z_V V_{\mu,{\rm Loc}}^a\,,\qquad A_{\mu,{\rm ren}}^a \, = \, Z_A A_{\mu,{\rm Loc}}^a\,,
\eea
and the renormalization factors $Z_{V},Z_{A}$ are fixed by imposing certain renormalization conditions, for example by imposing the continuum chiral Ward identities \cite{Karsten:1980wd,Bochicchio:1985xa,Maiani:1986yj}. For the vector current one may fix $Z_{V}$ using the forward matrix element \cite{Martinelli:1990ny}
\bea\label{RenCondV1}
Z_{V} \langle \pi^a({\bf p}) | V^{b}_{0,{\rm Loc}}(0) | \pi^c({\bf p})\rangle = 2 E\,.
\eea
Alternatively, since the conserved vector current does not get renormalized it may be used to obtain $Z_{V}$ according to \cite{Martinelli:1990ny,Martinelli:1993dq,Henty:1994ta}
\bea\label{RenCondV2}
Z_{V} =  \frac{\langle f | V^{a}_{\rm Con}(0) | i\rangle}{\langle f | V^{a}_{\rm Loc}(0) | i\rangle}\,.
\eea
In order to fix the renormalization of the axial vector the Ward identity 
\bea\label{RenAC}
\int_{\partial R} {\rm d}\sigma_{\mu}(x) \epsilon^{abc} \langle f |  A_{\mu, {\rm ren}}^{a}(x) A_{\nu,{\rm ren}}^{b}(y) |i\rangle = 2i \langle  f | V_{\nu,{\rm ren}}^{c}(y) | i \rangle
\eea
has been used \cite{Luscher:1996jn}, which is valid at vanishing PCAC quark mass. Since numerical simulations at $m_{{\rm PCAC}}$ are extremely difficult with standard periodic boundary conditions, the computation of $Z_{A}$ using  \pref{RenAC} is done by employing Schroedinger functional boundary conditions \cite{Jansen:1995ck,Luscher:1996jn}.

No matter which particular condition has been chosen for the lattice currents, the same condition must be imposed in the chiral effective theory in order to properly match the currents. 
Analogously to the lattice currents we introduce renormalized currents in the chiral effective theory according to \pref{RenCurrents} with renormalization constants $Z_{V}$ and $Z_{A}$. Suppose we fix $Z_{V}$ by either \pref{RenCondV1} or by \pref{RenCondV2}. Using the expressions for the currents we obtain 
\bea
Z_{V} & = & 1 + {\rm O} (a^{2}) \,.
\eea
Through O($a$) condition \pref{RenCondV2} gives a trivial $Z$-factor, simply because to this order both vector currents are identical. However, the 1 coming from condition \pref{RenCondV1} has a different origin and involves a cancellation of the O($a$) correction in the vector current and the wave function renormalization $Z_{\pi}$ stemming from the initial and final pion states. Consequently, using a two-pion state  -- with the appropriate quantum numbers and an appropriately modified right hand side -- condition \pref{RenCondV1} would yield  $Z_{V}  = Z_{\pi} \,=\, 1 - \frac{16}{f^{2}} a W_{45}\,.$
The difference of the two renormalization factors is of O($a$) and vanishes in the continuum limit, as expected. This simple example illustrates how the renormalization factor depends on the chosen renormalization condition. Consequently, the renormalized current and matrix elements of it will differ by terms of O($a$).

Note that the conditions \pref{RenCondV1} and \pref{RenCondV2} can be studied within the chiral effective theory since the hadronic states contain pions only.  Other conditions, involving a matrix element between the vacuum and a vector meson state, for example,  are not easily done in standard meson ChPT, since the vector meson is not a degree of freedom in the chiral effective theory. Conditions involving quark states \cite{Martinelli:1994ty} are even less tractable.
In practice, conditions involving pseudo scalar states only can be treated in the chiral effective theory. 

As an example for the renormalization constant $Z_{A}$ we impose the Ward identity \pref{RenAC} with single pion states. In the effective theory  there is no problem in setting $m_{PCAC}$ to zero, so for simplicity we assume an infinite volume, in contrast to Schroedinger functional BC. We obtain 
\bea\label{ZAeff}
Z_{A}  =  1 -\frac{8}{f^{2}} a \Big([2W_{A,1}+W_{A,2}] 
- W_{45}
\Big) + {\rm O}(a^{2})\,.
\eea 
Again, as for the vector current, we would have obtained a different result if we had imposed \pref{RenAC} with other than single pion states. 

\section{Illustration: The pion decay constant}

Having constructed and renormalized the currents we are ready to calculate matrix elements and observables. As an example we compute the matrix element of the axial vector current between the vacuum and a single pion state, giving the pion decay constant. 

Taking into account \pref{ZAeff} we find to one loop
\bea\label{eq:fpiNLO}
f_{\pi}  & = &  f\left(1 - \frac{1}{16\pi^{2}f^{2}}\left[1 + {a} {\cal C}_{1}\right] M^{2}_{\pi}\ln\frac{M^{2}_{\pi}}{\mu^{2}} + \frac{8}{f^{2}} M^{2}_{\pi} [L_{45} + {a} {\cal C}_{2}]\right)\,.
\eea
${\cal C}_{1},{\cal C}_{2}$ are combinations of the low-energy coefficients in the axial vector current. For $a\rightarrow 0$ we recover the familiar $N_{f}=2$ result of continuum ChPT. Away from the continuum limit the result is modified, but the leading corrections are of O($aM_{\pi}^{2}$), not of O($a$) as one might have expected. The reason for this 'improved' result is the use of the renormalized current and the renormalization condition \pref{RenAC}. In a way the leading O($a$) correction has been ``renormalized away''.  This is similar to the leading O($a$) correction to the pion mass, which contributes to the additive mass renormalization only \cite{Sharpe:1998xm}. 

Note that the combination $L_{45} = 2L_{4}+L_{5}$ of Gasser-Leutwyler coefficients appears in the lattice spacing dependent combination $L_{45}^{\rm eff}(a)=L_{45} + a {\cal C}_{2} $. This should be kept in mind if $L_{45}$ is determined at non-zero $a$ by monitoring the pion mass dependence of the decay constant. 

The coefficient of the chiral logarithm also receives  an O($a$) correction in the form of a factor $[1 + a {\cal C}_{1} ]$. Consequently, the coefficient of the chiral logarithm is, in contrast to continuum ChPT, not a universal coefficient depending on $N_{f}$ and $f$ only, but on the --non-universal-- lattice artifacts too. This fact has already been stressed in Ref.\ \cite{Aoki:2003yv} some time ago.
Hence, the coefficient of the chiral logarithm does not provide a stringent test for deciding whether the lattice simulation is performed in the chiral regime or not.  In view of  eq.\ \pref{eq:fpiNLO} it is not sufficient to reduce the quark masses only in order to see the expected continuum chiral logarithm, the continuum limit needs to be taken too.

Finally we want to comment on the origin of the discrepancies between our result \pref{eq:fpiNLO} and a previously published result for the decay constant in ref.\ \cite{Rupak:2002sm}. In this reference the Noether current, derived from the chiral Lagrangian, has been used for the axial vector current, which misses the contribution proportional to $W_{A,2}$ in \pref{SymACurrent}. Moreover, no explicit  renormalization condition has been taken into account. Consequently, the axial vector current does not satisfy the chiral Ward identity \pref{RenAC} and O($a$) corrections appear in the result for the decay constant which are absent in \pref{eq:fpiNLO}.

\section{Summary}

The matching of the local lattice vector and axial vector currents to their WChPT counterparts needs to take into account two aspects which are not present in continuum ChPT. Firstly, since the local currents are not conserved, they do not in general map onto the conserved currents in WChPT. In particular, the Noether procedure is not appropriate to get the effective currents. Secondly, the finite renormalization of the lattice currents at non-zero lattice spacing needs to be taken into account in order to properly match the currents.  This means that the same renormalization condition needs to be imposed in WChPT which has been chosen to fix the renormalization of the lattice currents. This implies non-trivial $Z$ factors in WChPT.

Depending on the particular choice for the renormalization conditions the expressions for the renormalized currents will differ by terms of O($a$). As a result the WChPT predictions for matrix elements of the currents will be different as well. Consequently, a result for an observable like $f_{\pi}$ should make reference to the renormalization condition one has chosen. This is expected, of course, since the same statement equally applies to the lattice data that WChPT is supposed to describe. 

\section*{Acknowledgments}

O. B.\  acknowledges useful discussions with Johan Bijnens, Maarten Golterman, Steve Sharpe and Rainer Sommer.

This work is supported in part by the Grants-in-Aid for
Scientific Research from the Ministry of Education,
Culture, Sports, Science and Technology
(Nos. 13135204, 15204015, 15540251, 16028201).



\end{document}